\documentclass{lmcs}
\pdfoutput=1
\usepackage[utf8]{inputenc}

\usepackage{lastpage}
\lmcsdoi{22}{1}{3}
\lmcsheading{}{\pageref{LastPage}}{}{}%
{Jun.~19,~2024}{Jan.~06,~2026}{}

\usepackage{tikz,bitset}
\usepackage{hyperref}

\usetikzlibrary{calc,decorations.pathreplacing,patterns}
\usetikzlibrary{arrows.meta, calc, decorations.pathmorphing}

\newcommand*{\nn}{\mathbb{N}}

\newcommand*{\im}{\mathrm{Im}}
\newcommand*{\spec}{\mathrm{Spec}}

\newcommand*{\SGR}[1]{\mathcal{G}_{#1}}

\newcommand*{\coNP}{{\ensuremath{\mathsf{coNP}}}}
\newcommand*{\NP}{{\ensuremath{\mathsf{NP}}}}

\newcommand*{\POLYTIME}{{\ensuremath{\mathsf{P}}}}

\newcommand*{\PSPACE}{{\ensuremath{\mathsf{PSPACE}}}}
\newcommand*{\NEXPTIME}{{\ensuremath{\mathsf{NEXPTIME}}}}

\newcommand*{\constant}[1]{\mathrm{constant}_{#1}}
\newcommand*{\constantl}[1]{\mathrm{constant}^\circ_{#1}}
\newcommand*{\recolor}[1]{\mathrm{recolor}_{#1}}
\newcommand*{\join}[1]{\mathrm{join}_{#1}}

\newtheorem{theoremintro}{Theorem}

\theoremstyle{thmC}
\newtheorem{theoremintroC}[theoremintro]{Theorem}

\begin{document}

\title[Hardness of MSO over succinct graphs]{Hardness of monadic second-order formulae\texorpdfstring{\\}{} over succinct graphs}
\thanks{
This works has received financial support from project
ANR-24-CE48-7504 ALARICE,
HORIZON-MSCA-2022-SE-01 101131549 ACANCOS,
STIC AmSud CAMA 22-STIC-02 (Campus France MEAE)}
\author[Gamard]{Guilhem Gamard\lmcsorcid{0000-0002-3951-9649}}[a,d,e,f]
\author[Goubault-Larrecq]{Ali\'enor Goubault-Larrecq}[b]
\author[Guillon]{Pierre Guillon\lmcsorcid{0000-0002-4665-6887}}[c]
\author[Ohlmann]{Pierre Ohlmann}[b]
\author[Perrot]{K\'evin Perrot}[b]
\author[Theyssier]{Guillaume Theyssier}[c]
\address{LORIA, Campus scientifique, BP 239, 54506, Vandoeuvre-lès-Nancy Cedex, France}
\email{guilhem.gamard@loria.fr}
\address{Aix Marseille Université, CNRS, LIS, Marseille, France} 
\email{alienor.goubault-larrecq@lis-lab.fr, pierre.ohlmann@lis-lab.fr, kevin.perrot@lis-lab.fr}
\address{Aix-Marseille Université, CNRS, I2M, Marseille, France}
\email{pierre.guillon@math.cnrs.fr, guillaume.theyssier@cnrs.fr}
\address{CNRS, Délégation Centre-Est, 17 rue Notre-Dame des Pauvres, BP 10075, 54519, Vandoeuvre-lès-Nancy Cedex, France}
\address{Université de Lorraine, 34 cours Léopold, 54052, Nancy Cedex, France}
\address{Inria MOCQUA, 615 Rue du Jardin-Botanique, 54600 Villers-lès-Nancy, France}

\begin{abstract}
  Our main result is a succinct \textit{counterpoint} to Courcelle's meta-theorem as follows: every 
  cw-nontrivial monadic second-order (MSO) property is either \NP-hard or \coNP-hard over graphs given by succinct representations.
  Succint representations are Boolean circuits computing the adjacency relation.
  Cw-nontrivial properties are those which have infinitely many models and infinitely many countermodels with bounded cliquewidth.

  Moreover, we explore what happens when the 
  cw-nontriviality condition is dropped and show that, under a reasonable complexity assumption, the previous dichotomy fails, even for questions expressible in first-order logic.
\end{abstract}
\maketitle

\section{Introduction}
  In this paper, we are interested in deciding properties of graphs defined in monadic second-order logic (MSO).
A series of results by Courcelle deals with this question; in particular~\cite{courcellebook} proves that every MSO property is decidable in linear time, given a graph with bounded treewidth (encoded by its adjacency matrix).
{The result also holds for the more general family of graphs with bounded cliquewidth~\cite{courcelle-cw}.}
Now what if the graph is not arbitrary, but presents some structure that allows a shorter encoding?
Assume that the graph is described succinctly, i.e, by a Boolean circuit which computes the adjacency relation between nodes, which are represented by binary numbers.
In this case the adjacency matrix might be exponentially larger than the circuit representation, so Courcelle's theorem does not give a polynomial-time algorithm.
One natural question is whether it is possible to exploit the circuits in better ways than just querying for all possible edges, in order to more directly deduce structural information about the graph.
Our main result essentially tells that it is impossible as soon as the property is nontrivial for bounded-cliquewidth graphs.
\begin{theoremintro}\label{t:succinct}
  If $\phi$ is a cw-nontrivial MSO sentence,
  then testing $\phi$ on graphs represented succinctly is either \NP- or \coNP-hard.
\end{theoremintro}
\emph{Cw-nontrivial} means that $\phi$ has infinitely many models with some fixed cliquewidth,
and infinitely many countermodels with some fixed cliquewidth.
Formal definitions appear in Section~\ref{sec:pre}, including a definition of \emph{cliquewidth}, and the formal version of this Theorem~\ref{t:succinct} is Theorem~\ref{thm:main}.
In this paper, we only consider \emph{finite models}.
Considering infinite models would require to agree on a computational model to describe them, say Turing machines, and the original Rice theorem then states that all nontrivial properties are even undecidable.

Succinct representations of graphs have already been considered in the literature.
In \cite[Table 1]{Galperin_1983}, the authors establish a list of natural graph properties that are polynomial-time solvable when using the usual input representations (matrix or adjacency list) but become (at least) \NP-hard when using succinct representations.
Interestingly, all of these properties fall into our meta-theorem~\ref{t:succinct}.
Moreover, their \cite[Theorem~3.1]{Galperin_1983} gives a sufficient condition for \NP-hardness, which can be compared to a weak form of our gluing lemmas, though our final statement, involving logic, is incomparable.
Finally, \cite[Open problem~1]{Galperin_1983} vaguely conjectures that all nontrivial properties are \NP-hard, provided that a good definition of nontriviality be given.
Theorem~\ref{t:succinct} proves this conjecture for MSO sentences and some natural notion of nontriviality (and adapting it by including the symmetric \coNP-hardness case), while Theorem~\ref{thm:bad} disproves it when this is relaxed, under reasonable complexity assumptions.

Later, \cite{Papadimitriou_1986} establishes that \NP-hard properties with the usual representation become \NEXPTIME-hard with succinct representations,
and~\cite{Balcazar_1992} exposes such complexity lower bound conversions for weaker classes,
proving for example that connectivity and planarity testing are \PSPACE-hard for succinctly represented graphs.
Although a complexity blowup is expected when taking the succinct version of a problem, this is not a general fact.
Actually,~\cite{DAS2017436} shows that when taking CNF or DNF formulae as succinct representations, there are examples of problems whose complexity does not increase when encoded in the new form, or increases to an intermediate complexity class less powerful than the exponential blow up. 
In the present paper, we focus on MSO model-checking on graphs with succinct representations by circuits.

Our motivation comes from the world of automata networks.
An \emph{automata network} can be seen as a computer network where all machines hold a local state and update synchronously by reading neighboring states and applying a local transition.
To update a machine (referred to as an \emph{automaton}) $v$ of the network, first collect the states of its inbound neighbors into a tuple,
and then feed that tuple as an input symbol to the update function of $v$.
Globally speaking, all automata are updated synchronously (though an extensive literature has explored other \emph{update modes}~\cite{ds20}),
so that the state of $v$ at time $t+1$ only depends on the states of its neighbors at time $t$.
One of the initial intents behind this definition was to model the dynamics of gene regulation~\cite{k69,t73,ma98,ks08}. 
Nowadays, automata networks are also used as a setup for distributed algorithms and as a modelling tool in engineering.
Those applications have motivated the study of automata network \emph{per se}
and many theoretical properties were found~\cite{e59,c71,r86,gm90,a08}. 

In general, the automata in a network may behave non-deterministically, so that the dynamics as a whole may be non-deterministic.
The local behavior of each automaton is typically described as a formula or relation,
which can be gathered into one Boolean circuit.
Given two binary words $z_1,\dots,z_n$ and $z'_1,\dots,z'_n$
(referred to as \emph{configurations}) assigning states to all automata of the network,
the circuit returns whether or not the first configuration can transition into the second one.
It succinctly encodes the directed graph
(referred to as the \emph{transition graph}) whose vertices are the configurations, and edges follow the transitions.
While seemingly artificial, this encoding is relevant for applications.
When automata networks are used to model actual computer networks, it is reasonable to assume access only to the source code of the programs run by the nodes.
Boolean circuits represent this source code.

The results from~\cite{ggpt21} already hint that the encoding cannot be smartly used to solve some questions efficiently. 
For instance, with deterministic automata networks encoded as $n$-bit-input $n$-bit-output circuits
(computing the unique successor of each configuration):
\begin{theoremintroC}[{\cite[Theorem 5.2]{ggpt21}}]
  \label{thm:fo}
  Let $\phi$ denote a question about graphs expressible in first-order logic.
  It is either $O(1)$, or \NP-hard, or \coNP-hard to decide, given a deterministic automata network as input,
  whether its transition graph satisfies $\phi$.
\end{theoremintroC}

In particular, first-order logic cannot express any nontrivial polynomial-time solvable question about the dynamics of deterministic automata networks, unless $\POLYTIME=\NP$.
This is a strong indication that it is indeed not tractable to analyze the Boolean circuits given to us in order to extract structural information about the dynamics: the best we can do is to evaluate the circuits to explore the transition graph.

\paragraph{Contributions.}

The present work started as an attempt to generalize Theorem~\ref{thm:fo} in two directions:
from first-order logic (FO) to monadic second-order logic (MSO),
and from deterministic automata networks to non-deterministic networks.
Neither generalization is trivial.

\emph{Questions about general automata networks are harder than questions about deterministic ones.}
For instance, the question ``is the automata network deterministic?'' is expressible in FO.
If we restrict ourselves to deterministic ones, that question is $O(1)$; in general, it is not.
Thence our generalizations prove, in particular, that determinism is either \NP- or \coNP-hard.

\emph{Questions in MSO are stronger than questions in FO.}
Properties expressible in MSO but not in FO include connectivity properties as well as minor-testing (therefore all minor-closed properties are expressible in MSO).
Thence our results prove, in particular, that testing whether the transition graph is series-parallel is either \NP- or \coNP-hard.

In addition to being technically harder, both generalizations are useful.
Indeed, when restricting ourselves to deterministic networks, we restrict ourselves to transition graphs of out-degree one.
This is a strong restriction from the perspective of succinct graphs taken in the present paper.
The generalization to non-deterministic networks lifts that restriction,
which enables future work to explore deeper connections between automata networks, Boolean circuits, and graph combinatorics.
Moreover, MSO logic allows to express the relation ``there is a chain of transitions from configuration $x$ to configuration $y$'',
which FO cannot.
That relation naturally arises in many practical questions.

In the non-deterministic setting, we do not get a general result: Theorem~\ref{t:succinct} requires the cw-nontriviality hypothesis.
Many of the properties considered in the literature so far, and in particular questions mentioned earlier in this introduction, are cw-nontrivial (up to turning counting questions into decision questions in the usual way,
e.g., ``how many fixpoints?'' becomes ``are there more than $k$ fixpoints?'' for a fixed $k$).
Examples of cw-trivial MSO-properties are ``being a square grid'', or ``being a square grid encoding a run of a given Turing machine''.
In fact, one can view our assumption of cw-nontriviality as a means of restricting to properties that are non-trivial over tree-like graphs, thereby excluding such complexity-theoretic properties.

The cw-nontrivial condition is crucial in our proofs, since it gives the existence of regular families of models and countermodels on which to build a polynomial reduction.
It is natural to ask whether it is necessary.
In Section~\ref{sec:bad} (Corollary~\ref{cor:bad}), we give a partial answer that we can state informally as follows:

\begin{theoremintro}
  \label{thm:noarb}
  There is a (cw-trivial) first-order sentence $\psi$ such that,
  under plausible complexity assumptions,
  testing $\psi$ on a given succinctly represented graph is neither
  constant time, nor \NP-hard, nor \coNP-hard.
\end{theoremintro}

To complete the picture around cw-triviality we establish that the decidability of the cw-nontrivial condition of our main Theorem depends on whether a bound on the cliquewidth is also given as input or not (propositions~\ref{prop:cw-triviality-undecidable} and \ref{prop:cw-triviality-decidable}).

\paragraph{Contents.}
 In Section~\ref{sec:pre}, we give the definitions and notations.
 In Section~\ref{sec:stmt}, we restate the main result and give a proof outline.
 In Sections~\ref{sec:saturating}--\ref{sec:sat}, we prove the main result;
 readers not interested the technical details can safely skip these sections.
 In Section~\ref{sec:bad}, we show the existence of counter-examples among cw-trivial sentences (under some complexity-theoretic hypothesis), and in Section~\ref{sec:deciding-cw-triviality} we discuss decidability of cw-triviality.
 We conclude with a discussion and suggestions for further research.

\section{Definitions}
\label{sec:pre}

\paragraph{Succinct graph representations.}
A (directed) graph $G$ is said to be \emph{succinctly represented} as a pair $(N,C)$,
where $C$ is a Boolean circuit on $2n$ inputs and one output and $N$ is an integer (encoded in binary) with ${N\leq 2^n}$,
whenever there is a one-to-one labeling of the vertices of $G$ onto ${\{0,\dots,N-1\}}$ such that ${C(x,y)=1}$ if and only if there is an edge from the vertex labeled $x$ to the vertex labeled $y$.
We denote $G=\SGR{N,C}$,
and we always assume that a Boolean circuit is at most polynomial in
the adjacency matrix of the encoded graph, because such an encoding always exists.
An example is given on Figure~\ref{fig:example}.

\begin{figure}
  \centering
  \begin{tikzpicture}
    \tikzstyle{conf} = [draw,rectangle,inner sep=1.5pt]
    \tikzstyle{gate} = [draw,circle,minimum size=12pt,inner sep=0pt]
    \tikzstyle{arc} = [-stealth]
    \foreach \c in {0,...,14}{
      \bitsetSetDec{bits}{\c}
      \ifthenelse{\equal{\bitsetGet{bits}{3}}{0}}{\def\y{0}\def\x{\bitsetGetDec{bits}-3.5}}{ 
        \ifthenelse{\equal{\bitsetGet{bits}{2}}{0}}{\def\y{1}\def\x{2*\bitsetGetDec{bits}-19}}{ 
          \ifthenelse{\equal{\bitsetGet{bits}{1}}{0}}{\def\y{2}\def\x{4*\bitsetGetDec{bits}-50}}{ 
            \def\y{3}\def\x{0} 
          }
        }
      }
      \node[conf] (n\bitsetGetBin{bits}{4}) at (\x,\y) {\bitsetGetBin{bits}{4}};
    }
    \foreach \c in {0,...,13}{
      \bitsetSetDec{bitsorigin}{\c}
      \bitsetSetDec{bits}{\c}
      \bitsetShiftRight{bits}{1} 
      \bitsetSet{bits}{3} 
      \draw[arc] (n\bitsetGetBin{bitsorigin}{4}) to (n\bitsetGetBin{bits}{4});
    }
    \begin{scope}[shift={(5,2)},xscale=.6]
      \draw[decorate, decoration = {brace}] (-.5,.3) -- node[above,yshift=2pt] {$x$} (3.5,.3);
      \draw[decorate, decoration = {brace}] (4.5,.3) -- node[above,yshift=2pt] {$y$} (8.5,.3);
      \foreach \x in {0,1,2,3,5,6,7,8}
        \node[gate] (i\x) at (\x,0) {};
      \foreach \x in {0,1,2}
        \node[gate] (e\x) at (2*\x+2,-1) {$=$};
      \foreach \x in {0,1}
        \node[gate] (a\x) at (2*\x+3,-1.5) {$\wedge$};
      \node[gate] (o0) at (4,-2) {$\wedge$};
      \draw[arc] (i0) to (e0);
      \draw[arc] (i1) to (e1);
      \draw[arc] (i2) to (e2);
      \draw[arc] (i6) to (e0);
      \draw[arc] (i7) to (e1);
      \draw[arc] (i8) to (e2);
      \draw[arc] (e0) to (a0);
      \draw[arc] (e1) to (a0);
      \draw[arc] (e2) to (a1);
      \draw[arc] (i5) to (a1);
      \draw[arc] (a0) to (o0);
      \draw[arc] (a1) to (o0);
    \end{scope}
  \end{tikzpicture}
  \caption{
    Example succinct representation of $G$ (left) as a Boolean circuit $C$ (right)
    on labels $\{0,\dots,14\}$, that is with $N=15$.
    Interpreting labels as binary numbers on 4 bits with the least significant bit on the right,
    there is an arc $(x,y)$ if and only if $y=\lfloor x/2 \rfloor+8$.
    We use a syntactic sugar ``$=$'' for Boolean equality $(a \wedge b) \vee (\neg a \wedge \neg b)$.
    Remark that generalizing this idea to construct a succinct representation of a binary tree on $2^n-1$
    nodes requires circuits of size linear in $n$ (here $n=4$).
  }
  \label{fig:example}
\end{figure}

\paragraph{MSO Logic.}
A \emph{$k$-colored graph} $G = (V, E, C_G)$ 
is a graph where each vertex is assigned one color by $C_G:V\to  \{1,\dots,k\}$.
We consider MSO formulae over $k$-colored graphs.
These have two kinds of variables corresponding to vertices ($x_1, x_2, \dots$) and sets of vertices ($X_1, X_2, \dots$)---
Accordingly, there are two kinds of existential quantifiers: existence of a vertex, and existence of a set of vertices.
The Boolean connectives $\neg$ and $\wedge$ are as usual.
The universal quantifiers, bounded quantifiers and other Boolean connectives are derived from them.
The atoms are: equality ($x_1=x_2$),
membership ($x_1\in X_1$),
colors ($C_G(x_1)=i$ for $i \in \{1,\dots,k\}$)
and adjacency relation $E(x_1,x_2)$, meaning that $\SGR{N,C}$ has an edge from $x_1$ to $x_2$.
Note that the adjacency symbol $E$ is a relational symbol in our signature, \emph{not} a functional symbol.

Given a $k$-colored graph $G$, together with vertices $a_1,\dots,a_n$ and subsets of vertices $A_1,\dots,A_m$, and a formula $\phi(x_1,\dots,x_n,X_1,\dots,X_m)$, we write $(G,a_1,\dots,a_n,X_1,\dots,X_m) \models \phi(x_1,\dots,x_n,X_1,\dots,X_m)$ if $\phi$ is satisfied, with the natural semantics, when assigning variables in the obvious way.
A \emph{sentence} is a closed monadic second-order formula, i.e., one where all variables are bound to a quantifier.
Here are some examples, 
\begin{itemize}
\item \emph{(Existence of a loop)} $\exists x: E(x,x)$.
\item \emph{(Unicity of the loop)} $\forall x, \forall x': E(x,x) \wedge E(x',x') \implies x=x'$.
\item \emph{(Determinism)} $\forall x, \forall y, \forall y': E(x,y) \wedge E(x,y')\implies y=y'$.
\item \emph{(Nontrivial cycle)} $\exists X,[\exists x\in X]\wedge[\forall x\in X,\exists y\in X:x\neq y\wedge E(x,y)]$.
\end{itemize}
With this MSO signature, it is possible to express as a macro the relation $E^*(x,y)$: ``there exists a chain $x=z_1,z_2,\dots,z_{n-1},z_n=y$ such that $E(z_{k},z_{k+1})$ holds for every $1\leq k\leq n-1$.''
Indeed, this amounts to the sentence $\forall X,[x\in X\wedge\forall z\in X,\forall z',E(z,z')\implies z'\in X]\implies y\in X$.
On the other hand, it is not possible to express something like: ``all configurations have the same out-degree.''
When a graph $G$ satisfies a sentence $\phi$, we call it a \emph{model} and denote and call it a \emph{countermodel} otherwise.
The \emph{quantifier rank} of a formula is its number of nested quantifiers (regardless of their alternation).
See~\cite{l04} or~\cite{df13} for more information about MSO logic on graphs.

\paragraph*{Types.}
Let $G$ be a $k$-colored graph, $a_1,\dots,a_n$ be vertices of $G$ and $A_1,\dots,A_m$ be subsets of vertices of $G$.
The \emph{quantifier rank $q$ type of $(G,a_1,\dots,a_n,A_1,\dots,A_m)$} is the set of all MSO-formulae with $n$ free point variables and $m$ free set variables and quantifier rank $\leq q$ that are satisfied by $(G,a_1,\dots,a_n,A_1,\dots,A_m)$.
We say that a set of formulae is a \emph{realized $(n,m)$-type of quantifier rank $q$} if it is a type as above.
The following standard result follows from an easy induction on $q$.

\begin{lemC}[{\cite[Proposition~7.5]{l04}}]\label{lemma:finitely-many-types}
  Given $n,N$ and $q$, there are finitely many realized $(n,m)$-types of quantifier rank $q$.
\end{lemC}

We write $(G,a_1,\dots,a_n,A_1,\dots,A_m)\equiv_q(H,b_1,\dots,b_n,B_1,\dots,B_m)$ if the corresponding types are the same.
By a slight abuse, we simply say ``types'' to refer to $(0,0)$-types (i.e., sets of sentences).
We let $T_q$ denote the (finite) set of realized types of quantifier rank $q$ (the fixed number $k$ of colors will always be clear from the context).

\paragraph{Clique decompositions and cliquewidth.}

Let $\oplus$ denote the disjoint union between sets (possibly renaming some elements if they were not literally disjoint).
Consider the following four operations used
to construct $k$-colored graphs:
\begin{enumerate}[label=(\roman*)]
\item $\constant i$ (arity 0): given a color $i \in C$ and a vertex label $v$, it returns the graph 
    $G = (\{v\}, \emptyset, C_G)$ with $C_G(v) = i$;
  \item constant$^\circ_i$ (arity 0): given a color $i \in C$ and a vertex label $v$, it returns the graph 
    $G = (\{v\}, \{(v,v)\}, C_G)$ with $C_G(v) = i$;
  \item $\recolor f$ (arity 1): for a \emph{recoloring} function $f : C \to C$ and given a $k$-colored graph $H$, it returns 
    the graph $G = (V(H), E(H), C_G)$ such that $C_G = f \circ C_H$;
  \item $\join M$ (arity 2): for a set
    $M \subseteq \{(i, j, d) \mid i \in C, j \in C, d \in \{R,L\}\}$
    and given two $k$-colored graphs $H$ and $H'$, it
    returns the graph $G = (V(H) \oplus V(H'), E(H) \oplus E(H') \oplus E_R \oplus E_L, C_G)$, where 
    $E_R = \{(v_i, v_j) \mid (i,j,d)\in M, d=R, v_i \in H, C_H(v_i) = i, v_j \in H',C_{H'}(v_j) = j\}$
    which are the arrows going from
    vertices of color $i$ in $H$ to vertices of color $j$ in $H'$ for triples with $d=R$ (right, from $H$ to $H'$),
    and symmetrically for
    $E_L = \{(v_i, v_j) \mid (i,j,d)\in M, d=L, v_i \in H', C_{H'}(v_i) = i, v_j \in H,C_H(v_j) = j\}$
    with $d=L$ (left, from $H'$ to $H$).
    Moreover, $C_G(v) = C_H(v)$ if $v \in H$ and
    $C_G(v) = C_{H'}(v)$ otherwise.
\end{enumerate}

A clique decomposition is a term in the free algebra generated by the above operations.
Stated differently, it is a rooted tree where leaves are labeled by $\constant i$ or $\constantl i$ for some $i$, unary nodes are labeled by $\recolor f$ for some $f$, binary nodes are labeled by $\join M$ for some $M$, and there are no higher-degree nodes.
A clique decomposition $\mathcal C$ generates a $k$-colored graph in the natural way.

We say that
$\mathcal{C}_G$ is a clique decomposition of a (non-colored) graph $G$ if there exists a coloring $C_G$
of the graph $G$ such that $\mathcal{C}_G$ is a clique decomposition of
$(V(G), E(G), C_G)$.
A graph may have several clique decompositions (be it colored or not), but a
clique decomposition corresponds to only one colored graph.
Moreover, one colored graph has exactly one corresponding (non-colored) graph.
Hence, by abuse we will say for a formula $\phi$ that a clique decomposition
$\mathcal{C}$ is a model (resp.~countermodel) of $\phi$ if $\mathcal C$ constructs a graph $G$ which is a
model (resp.~countermodel) of $\phi$.

The \emph{width} of a decomposition is the number $k$ of colors used in the decomposition.
The \emph{cliquewidth} of a graph is the minimal width among all of its clique decompositions. For 
every integer $k$, a $k$-clique decomposition means a clique decomposition of width at most $k$.
An example of clique decomposition is illustrated in Figure~\ref{fig:cliquedecomposition}.

\begin{figure}
  \centering
  \begin{tikzpicture}
    \tikzstyle{onode} = [draw,rectangle,rounded corners]
    \tikzstyle{vert} = [draw=black,circle,minimum size=10pt,inner sep=0pt]
    \tikzstyle{arc} = [-stealth]
    \node[onode] (r) at (0,0) {
      \begin{tikzpicture}
        \foreach [count=\x from 0] \color in {red,blue,green}
          \foreach \y in {0,1}
            \node[vert,fill=\color] (n\x\y) at (\x,\y) {};
        \draw[arc,loop left] (n00) to (n00);
        \draw[arc,loop left] (n01) to (n01);
        \draw[arc] (n10) to (n00);
        \draw[arc] (n11) to (n01);
        \draw[arc] (n20) to (n10);
        \draw[arc] (n20) to (n11);
        \draw[arc] (n21) to (n10);
        \draw[arc] (n21) to (n11);
        \draw[arc,bend left] (n20) to (n21);
        \draw[arc,bend left] (n21) to (n20);
      \end{tikzpicture}
    };
    \node[onode] (r0) at (-2,-2) {
      \begin{tikzpicture}
        \foreach [count=\x from 0] \color in {red,blue}
          \foreach \y in {0,1}
            \node[vert,fill=\color] (n\x\y) at (\x,\y) {};
        \draw[arc,loop left] (n00) to (n00);
        \draw[arc,loop left] (n01) to (n01);
        \draw[arc] (n10) to (n00);
        \draw[arc] (n11) to (n01);
      \end{tikzpicture}
    };
    \draw (r0) -- (r);
    \node[onode] (r1) at (2,-2) {
      \begin{tikzpicture}
        \foreach \y in {0,1}
          \node[vert,fill=green] (n2\y) at (2,\y) {};
        \draw[arc,bend left] (n20) to (n21);
        \draw[arc,bend left] (n21) to (n20);
      \end{tikzpicture}
    };
    \draw (r1) -- (r);
    \node[onode] (r00) at (-3.2,-3.5) {
      \begin{tikzpicture}
        \foreach [count=\x from 0] \color in {red,blue}
          \node[vert,fill=\color] (n\x0) at (\x,0) {};
        \draw[arc,loop left] (n00) to (n00);
        \draw[arc] (n10) to (n00);
      \end{tikzpicture}
    };
    \draw (r00) -- (r0);
    \node[onode] (r01) at (-0.8,-3.5) {
      \begin{tikzpicture}
        \foreach [count=\x from 0] \color in {red,blue}
          \node[vert,fill=\color] (n\x0) at (\x,0) {};
        \draw[arc,loop left] (n00) to (n00);
        \draw[arc] (n10) to (n00);
      \end{tikzpicture}
    };
    \draw (r01) -- (r0);
    \node[onode] (r10) at (2,-3.5) {
      \begin{tikzpicture}
        \foreach \x in {0,1}
          \node[vert,fill=blue] (n\x) at (\x,0) {};
        \draw[arc,bend left] (n0) to (n1);
        \draw[arc,bend left] (n1) to (n0);
      \end{tikzpicture}
    };
    \draw (r10) -- (r1);
    \newcommand{\constantnode}[1]{
      \begin{tikzpicture}
        \node[vert,fill=#1] {};
      \end{tikzpicture}
    }
    \newcommand{\constantloopnode}[1]{
      \begin{tikzpicture}
        \node[vert,fill=#1] (n) {};
        \draw[arc,loop left] (n) to (n);
      \end{tikzpicture}
    }
    \foreach \x in {0,1}{
      \node[onode] (r0\x0) at (-3.2-.6+\x*2.4,-5) {\constantloopnode{red}};
      \draw (r0\x0) -- (r0\x);
      \node[onode] (r0\x1) at (-3.2+.6+\x*2.4,-5) {\constantnode{blue}};
      \draw (r0\x1) -- (r0\x);
    }
    \foreach \x in {0,1}{
      \node[onode] (r10\x) at (2-.5+\x*1,-5) {\constantnode{blue}};
      \draw (r10\x) -- (r10);
    }
  \end{tikzpicture}
  \caption{
    Example of clique decomposition, with colors
    $C=\{0,1,2\}$ where $0$ is red, $1$ is blue, $2$ is green.
    Each node contains the corresponding $k$-colored graph,
    and the operation should easily be deduced from the arity.
    As an example, the join operation on the bottom right leading to a directed cycle of length two
    with blue nodes has $M=\{(1,1,R),(1,1,L)\}$.
  }
  \label{fig:cliquedecomposition}
\end{figure}

For two
clique decompositions $\mathcal{C}$ and $\mathcal{C'}$, we write  
$\mathcal{C} \oplus \mathcal{C'}$ for the clique decomposition 
$\textnormal{join}_{\emptyset}(\mathcal{C}, \mathcal{C'})$.

\paragraph{Remark.}
While being slightly non-standard, our definition has the useful property that to decide if an edge $(v,v')$ exists, for $v \neq v'$, it suffices to consider the least common ancestor of $v$ and $v'$ in the decomposition, which is necessarily a $\join M$ node, and the composition of the recolorings on the paths from $v$ and $v'$ to this node.
This will simplify our construction, and creates no increase in the cliquewidth
compared to the classical definition of clique decomposition for directed graphs using
constant nodes, disjoint unions, recolorings and edge insertions
(like in~\cite[Section~2.2]{frrs09} for example).

\paragraph*{Compositionality.} The choice of the operations above is motivated by the following standard result which is usually called the compositionality lemma.
It is the key result in Courcelle's celebrated theorem.

\begin{lemC}[{\cite[Corollary 5.60]{courcellebook}}]\label{lem:compositionality}
  Let $q \in \nn$.
  For the four operations above, the quantifier rank $q$ type of the output depends only on the quantifier rank $q$ types of the inputs.
\end{lemC}

Stated differently, for each operation above, say of arity $r$ (so $r \in \{0,1,2\}$), we have a table $T_q^r \to T_q$, where $T_q$ is the set of quantifier rank $q$ types, which gives the output type for the operation depending on the type of its inputs.
Together, these tables can be viewed as a deterministic bottom-up tree automaton, with state-space $T_q$, which reads clique decompositions and computes their quantifier rank $q$ type.

Note that as an important special case of Lemma~\ref{lem:compositionality}, which corresponds to the operator $\join \emptyset$, we get that the type of a disjoint union depends only on the types of its components.

\paragraph{Additional conventions.}
If $G$ is a graph, let $|G|$ denote the number of its nodes, dubbed its \emph{size}.
If $S$ is an instance of SAT, let $|S|$ denote the number of its variables, also dubbed its \emph{size}.
Unless stated otherwise: \emph{increasing} means strictly increasing;
\emph{integer} means positive or zero integer;
\emph{polynomial} means nonconstant polynomial with integer coefficients.

\section{Statement of the main result and proof outline}
\label{sec:stmt}

\paragraph{The problem.}
Given an MSO sentence $\phi$, 
define the model checking problem of $\phi$ on graphs given by a succinct representation.
\begin{enumerate}[label=]
\item \textsc{Succinct-}$\phi$
\item \emph{Input:} a succinct graph representation ${(N,C)}$. 
\item \emph{Output:} does $\SGR{N,C}\models\phi$?
\end{enumerate}
Observe that $\phi$ is \emph{not} part of the input: it is considered constant.
In other words, we have a family of problems parameterized by MSO sentences.

Our main result is a counterpoint to Courcelle's theorem:

\begin{thm}
  \label{thm:main}
  If there exists $k$ such that $\phi$ has infinitely many models and infinitely many countermodels with cliquewidth $\leq k$,
  then \textsc{Succinct-}$\phi$ is either \NP-hard or \coNP-hard.
  {The result still holds if we restrict inputs to graphs with bounded cliquewidth (cliquewidth at most $k$, for large enough $k$ depending on $\phi$).
    }
\end{thm}

An MSO sentence is 
\emph{cw-nontrivial} if it satisfies the assumption of Theorem~\ref{thm:main},
and it is \emph{cw-trivial} otherwise.
All examples of sentences given in the previous section are cw-nontrivial.
We will discuss sentences that are cw-trivial in Section~\ref{sec:bad}.

\paragraph{Proof outline.}
First, we show that there exists a ``good'' graph, call it $\Omega$,
such that $\Omega\oplus G$ is always a model of $\phi$ (where $\oplus$ denotes disjoint union),
no matter what $G$ is.
Then, we show that there exists a ``bad'' graph, say $Y$,
such that $Y\oplus\dots\oplus Y$ is always a countermodel of $\phi$, no matter how many disjoint copies of $Y$ we put.
We can arrange things so that $\Omega$ and $Y$ have the same number of vertices.

Now we perform a reduction: we are given an instance $S$ of SAT with $s$ Boolean variables,
and we produce a succinct graph representation $(N,C)$ such that $\SGR{N,C}\models\phi$ if and only if $S$ has at least one positive assignment.
We take $N=2^s\cdot|\Omega|=2^s\cdot|Y|$ so that we have $2^s$ groups of $|\Omega|=|Y|$ vertex labels.
For each vertex label ($n$-bit string), the circuit $C$ interprets the first $s$ bits as an assignment of the variables of $S$,
and evaluates $S$ on that assignment.
If it finds ``true'', then the corresponding $|\Omega|$ vertex labels realize a copy of $\Omega$ in $\SGR{N,C}$.
If it finds ``false'', then the corresponding $|Y|$ vertex labels realize a copy of $Y$ instead.
Consequently, $\SGR{N,C}$ contains as many copies of $\Omega$ as positive assignments for $S$,
and as many copies of $Y$ as negative assignments for $S$.
This completes the reduction: if there is at least one positive assignment, the defining property of $\Omega$ guarantees that the graph satisfies $\phi$.
Otherwise, the graph is only a pack of disjoint copies of $Y$, which does not satisfy $\phi$.

This whole construction can be performed in polynomial time because $\Omega$ and $Y$ do not depend on $S$:
they only depend on $\phi$, hence they are of constant size
(recall that $\phi$ is \emph{not} part of the input of the problem).
The only part of $C$ that depends on $S$ is the evaluation of $S$
which is easy to implement as a Boolean circuit.

Of course, things are not that simple.
First problem: we are actually unable to control whether $\Omega$ (called a \emph{saturating graph}) turns every graph into a model or every graph into a countermodel.
In the latter case, 
we will have to symmetrize all the remainder of the proof (in particular, $Y\oplus\dots\oplus Y$ will have to be a model),
and in that case we will get \coNP-hardness instead of \NP-hardness.
On the other hand, perhaps surprisingly, it turns out that $\Omega$ does not depend on $\phi$, but only on its the quantifier rank. 
It may be big but, being constant, its clique-width is also treated as a constant.
The details are explained in Section~\ref{sec:saturating}.


Second problem: we have to relax the requirements on $Y$.
What we will actually get is a triple of clique decompositions $(\mathcal{X},\mathcal{Y},\mathcal{Z})$ such that the graph constructed by
$\mathcal{X}\triangleleft \mathcal{Y}\triangleleft \dots\triangleleft \mathcal{Y}\triangleleft \mathcal{Z}$ is a countermodel of $\phi$ (or, if needed, a model of $\phi$), no matter how many copies of $\mathcal{Y}$ are in there.
The gluing operator $\triangleleft$ is more general than disjoint union; $\mathcal{C}\triangleleft \mathcal{C'}$ basically means:
``plug the root of clique decomposition $\mathcal{C'}$ at the marked node of 
clique decomposition $\mathcal{C}$''
(hence the $k$-colored graph constructed by $\mathcal{C'}$ somehow replaces some predefined leaf of $\mathcal{C}$).
The details are explained in Section~\ref{sec:pump}
(for notational convenience, $\mathcal{X},\mathcal{Y},\mathcal{Z}$ are called $\mathcal{C}_1,\mathcal{C}_2,\mathcal{C}_3$ in that section---the subscripts come in handy).


Third problem: the above reduction from SAT should be adapted following the concessions just made on $Y$.
Indeed, a priori, deciding whether a couple of vertices are adjacent may require to parse an exponentially long clique decomposition.
Fortunately, the clique decomposition is very regular, which allows to overcome this difficulty by using idempotence of recolor nodes in $\mathcal{Y}$ and
to compute adjacencies with a small circuit.
The details are explained in Section~\ref{sec:sat}, which also includes the final proof of Theorem~\ref{thm:main}.

Sections~\ref{sec:saturating}, \ref{sec:pump} and~\ref{sec:sat} each start with a proposition, and the remainder of the section is the proof of the proposition.
These three propositions together quickly yield a proof for Theorem~\ref{thm:main}.

\paragraph*{Comparison with~\cite{ggpt21}.}
In~\cite{ggpt21}, a similar proof outline is used to establish the main result about deterministic automata networks (see Theorem~\ref{thm:fo} in the introduction).
However, constructions in~\cite{ggpt21} for the ``good'' and ``bad'' graphs are somewhat specific and a bit tedious, we now employ more powerful tools from finite model theory.

\section{A graph saturating all sentences of fixed quantifier rank}
\label{sec:saturating}

The goal of this section is to construct the ``good'' graph $\Omega$ from the overview above.
It should satisfy the requirements from the following proposition.

\begin{prop}
  \label{pro:saturating}
  Fix $k,q\in\nn$.
  There exists a $k$-colored graph $\Omega_q$ such that,
  for every MSO sentence $\phi$ of rank $q$, either:
  \begin{enumerate}[label=(\roman*)]
  \item for every graph $G$, we have $G\oplus\Omega_q\models\phi$; or
  \item for every graph $G$, we have $G\oplus\Omega_q\not\models\phi$.
  \end{enumerate}
\end{prop}

In the first case, we say that $\Omega_q$ is a \emph{sufficient subgraph} for $\phi$;
in the second case, $\Omega_q$ is a \emph{forbidden subgraph} for $\phi$.
A graph that is either sufficient or forbidden for a given sentence is called \emph{saturating} for that sentence.
The rest of this section is a proof of Proposition~\ref{pro:saturating}.

Write $G\oplus G'$ for the disjoint union of a copy of $G$ and a copy of $G'$.
If $\ell$ is an integer, write $\bigoplus^\ell G$ the disjoint union of $\ell$ copies of $G$.

\begin{lem}
  \label{lem:qexists}
  For every integer $q$ and every nonempty $k$-coloured graph $G$,
  there exists an integer $p(G,q)$ such that for all $\ell,\ell' \geq p(G,q)$ it holds that $\bigoplus^{\ell} G\equiv_q \bigoplus^{\ell'} G$.
\end{lem}

\begin{proof}
  We show the lemma by induction on $q$.
  For $q=0$, we may set $p(G,0)=1$ and then there is nothing to prove since there is only one possible quantifier rank $0$ type with no free variables: the empty set.
  We let $q>0$ and assume the result proved for $q-1$ and any graph $G'$.
  
  We let $p'$ be the maximal value of $p(G',q-1)$, where $G'$ ranges over all possible $2k$-colored graphs obtained from $G$ by splitting each color class into two fresh ones.
  Then we set $p(G,q) = 2^{|G|} p'$.

  Let $\ell,\ell' \geq p(G,q)$ and let $H = \bigoplus^\ell G$ and $H'= \bigoplus^{\ell'} G$. 
  Consider a sentence $\phi$ such that $H \models \phi$; we should prove that also $H' \models \phi$.
  Without loss of generality we may assume that $\phi$ is either of the form $\exists x \psi(x)$ or $\exists X \psi(X)$.
  This is because the proofs are trivial for Boolean connectives.
  We treat the two cases separately.

  \paragraph*{Case 1: $\phi = \exists x \psi(x)$.}
  Since $H \models \phi$, there is $a \in H$ such that $(H,a)\models \psi(x)$.
  Now $a$ can be seen as a vertex of $G$.
  Consider the $(k+1)$-colored graph $G'$ obtained from $G$ by giving color $k+1$ to vertex $a$.
  Since $p(G,q) \geq p(G,q-1) +1$,
  it follows from the induction hypothesis that $\bigoplus^{\ell-1} G \equiv_{q-1} \bigoplus^{\ell'-1} G$,
  and therefore by compositionality (Lemma~\ref{lem:compositionality}), we get:
  \[
    G' \oplus \bigoplus^{\ell-1} G \equiv_{q-1} G' \oplus \bigoplus^{\ell'-1} G.
  \]
  By a straightforward translation between formulae involving graphs with a single vertex of color $k+1$ (such as above), and graphs with an identified vertex, we deduce that $(H,a)\equiv_{q-1}(H',a)$, and thus $H' \models \phi$.

  \paragraph*{Case 2: $\phi = \exists X \psi(X)$.}
  Since $H \models \phi$, there is $A \subseteq H$ such that $(H,A)\models \psi(X)$.
  Now $A$ can be seen as a map $f_A:2^G \to \nn$ assigning to each subset $S$ of $G$ the number of occurrences of $S$ as the intersection between $A$ and a copy of $G$ in $H$.
  Note that fixing a subset $S$ of $G$ is just the same as cutting each color class in two parts; more precisely, there is a bijection between quantifier rank $q-1$ types of $(G,S)$ and those of the $2k$-colored graph $G'$ obtained from $G$ by splitting each color graph into their intersections with $S$ and with its complement.

  It is not hard to see that thanks to the large enough value of $\ell'$, we may choose $B \subseteq H'$ in such a way that the associated map $f_B$ is so that $f_B(S)=f_A(S)$ if $f_A(S)< p'$ and $f_B(S) \geq p'$ if $f_A(S)\geq p'$. 
  Following the above discussion, the induction hypothesis yields that for each $S$, $H_S \equiv_{q-1} H'_S$, where $H_S$ (resp. $H'_S$) denotes the disjoint unions of all copies of $G$ in $H$ (resp. in $H'$) whose intersection with $A$ (resp. with $B$) matches $S$.

  We conclude by compositionality that $(H,A)\equiv_{q-1}(H',B)$ and in particular, $(H',B) \models \psi(X)$ and thus $H' \models \phi$.
\end{proof}

We are now ready to prove the proposition.

\begin{proof}[Proof of Proposition~\ref{pro:saturating}.]

  Call $a(q)$ the number of different realized quantifier rank $q$ types (it is finite by Lemma~\ref{lemma:finitely-many-types}), and let $G_1,\dots,G_{a(q)}$ denote some arbitrarily chosen graphs which realize these types.

  Define $\Omega_q$ as follows: take $p(G_i,q)$ disjoint copies of $G_i$, for $i$ ranging in $\{1,\dots,a(q)\}$.
  We show that $\Omega_q$ is either a forbidden or a sufficient subgraph for $\phi$.

  For every graph $G$, we have $G \equiv_q G_i$ for some $i$ in $\{1,\dots,a(q)\}$.
  Hence by compositionality, $\Omega_q \oplus G \equiv_q \Omega_q \oplus G_i$.
  However, since there are already $p(G_i,q)$ copies of $G_i$ in $\Omega_q$, and again applying compositionality, we get that $\Omega_q \oplus G_i \equiv_q \Omega_q$ thanks to Lemma~\ref{lem:qexists}.
  Thus $\Omega_q\equiv_{q} \Omega_q\oplus G$, and whether it is a model of $\phi$ or not does not depend on $G$.
\end{proof}

\section{Pumping models of cw-nontrivial sentences}
\label{sec:pump}

The goal of this section is to construct the ``bad'' graph from the proof overview.
As explained previously, it is in fact a triple of clique decompositions that are assembled in a linear fashion: a fixed prefix ($\mathcal C_1$ below), a pumpable infix ($\mathcal C_2$ below) and a fixed suffix ($\mathcal C_3$ below).
To formalize this, we need a way to compose clique decompositions, which we introduce now.
Throughout the section, we fix an integer $k$, and assume that all considered clique decompositions have width $\leq k$ and thus generate $k$-colored graphs.

%

\newcommand{\C}{\mathcal C}

A \emph{marked clique decomposition} is a term in the free algebra generated by the previous operations ($\constant i, \constantl i, \recolor f$ and $\join M$), plus an additional 0-ary operator $\square$ which acts as a placeholder.
We moreover assume that $\square$ appears exactly once.
Stated differently, it is the same as a clique decomposition, except that a unique leaf is labeled by $\square$ instead of $\constant i$ or $\constantl i$.

For a clique decomposition $\C$ which may or may not be marked, and a marked clique decomposition $\C'$, we let $\C \triangleleft \C'$ denote the clique decomposition obtained by substituting the placeholder $\square$ in $\C'$ with $\C$ (the root in the place of the $\square$).
Note that $\C \triangleleft \C'$ is marked if and only if $\C$ is marked.
This operation is associative so we may write $\C_1 \triangleleft \C_2 \triangleleft \dots \triangleleft \C_n$.



As a simple consequence of compositionality (Lemma~\ref{lem:compositionality}), we get the following statement.

\begin{lem}\label{lem:compositionality-oplus}
  For every clique decomposition $\C$ of width $k$, and every integer $q$, there is a map $\Delta_\C:T_q \to T_q$, where $T_q$ denotes the set of realizable types of quantifier rank $q$ (for $k$-colored graphs), such that for every clique decomposition $\C'$, the type of the graph generated by $\C' \triangleleft \C$ is obtained from the type of the graph generated by $\C'$ by applying $\Delta_\C$.
\end{lem}

We are now ready to prove the following proposition, which collects the requirements on the sought ``bad'' graph.

\begin{prop}
	\label{pro:pump}
	Let $\phi$ be an MSO sentence and $k$ an integer.
	If $\phi$ has infinitely many models of cliquewidth at most $k$,
	then there exist clique decompositions
        $\mathcal{C}_1,\mathcal{C}_2,\mathcal{C}_3$
        such that
        $
          \C_1 \triangleleft \C_2 \triangleleft \dots \triangleleft \C_2 \triangleleft \C_3
        $
        is a model of $\phi$, whatever the number of occurrences of $\C_2$.
        Furthermore, $\mathcal{C}_2$ contains at least one constant node (i.e., it is not the trivial marked clique decomposition comprised of a unique node labelled by $\square$).
\end{prop}

The proposition follows from applying the standard pumping lemma for tree-automata (see for instance~\cite[Section 1.2]{comon08}) to the automaton alluded to in the preliminaries.
For completeness, we now present a proof avoiding the automata-theoretic terminology.

\begin{proof}
  \newcommand{\op}{\mathrm{op}}
  
  Since $\phi$ has infinitely many models of cliquewidth at most $k$, and since there are finitely many types, there must be a type $t_0 \in T_q$ which contains $\phi$ and is realized by infinitely many graphs of cliquewidth at most $k$.
  Therefore, there is a clique decomposition $\C$ of height $>|T_q|$ which generates a graph of type $t_0$.
  For each node of $\C$, we consider the clique decomposition obtained by restricting $\C$ to descendants of this node, and label the node by the type of the corresponding graph.
  By the pigeonhole principle and our assumption on the depth of $\C$, there is a root-to-leaf path in $\C$ such that the above labels has a repetition, say that type $t$ has two occurrences on this path corresponding to nodes $u$ and $u'$ such that $u'$ is a strict descendant of $u$. 

  Then we may split $\C$ accordingly (see Figure~\ref{fig:context}): 
  \begin{itemize}
    \item $\C_1$ is obtained by restricting to descendants of $u'$;
    \item $\C_2$ is obtained by restricting to descendants of $u$, then substituting $u'$ with $\square$ (and removing its descendants, if any);
    \item $\C_3$ is obtained from $\C$ by substituting $u$ with $\square$ (and removing its descendants).
  \end{itemize}
Note that $\C_2$ is nontrivial since $u \neq u'$.
Consider the map $\Delta_{\C_2}$ from Lemma~\ref{lem:compositionality-oplus}.
Note that the label of $u'$ correspond to the type of the graph generated by $\C_1$, and the type of $u'$ corresponds to the graph generated by $\C_1 \triangleleft \C_2$.
By our assumption that both these types are equal to $t$, it follows that $\Delta_{\C_2}(t)=t$.

\begin{figure}
  \centering
  \begin{tikzpicture}[scale=.55,baseline]
    \foreach \s [count=\i] in {0,1.5,3} {
      \coordinate (a\i) at ($(0,0)  +(0,-\s)$);
      \coordinate (b\i) at ($(2,3)  +(0,-\s)$);
      \coordinate (c\i) at ($(4,0)  +(0,-\s)$);
      \coordinate (d\i) at ($(3,0)  +(0,-\s)$);
      \coordinate (e\i) at ($(2,1.5)+(0,-\s)$);
      \coordinate (f\i) at ($(1,0)  +(0,-\s)$);
    }
    \draw (a3)--(f2)--(a2)--(f1)--(a1)--(b1)--(c1)--(d1)--(c2)--(d2)--(c3)--cycle;
    \node[left,xshift=-6pt] at (b1) {$\mathcal{C}$};
    \draw (b2) node{$\bullet$} node[right]{$u$};
    \draw (b3) node{$\bullet$} node[right]{$u'$};
    \draw[decorate, decoration={random steps,segment length=5pt,amplitude=2pt}] (b1) -- (b2) -- (b3) -- (f3);
  \end{tikzpicture}
  \qquad
  $\longrightarrow$
  \qquad
  \begin{tikzpicture}[scale=.55,baseline]
    \foreach \s [count=\i] in {0,1.5,3} {
      \coordinate (a\i) at ($(0,0)  +(0,-\s)$);
      \coordinate (b\i) at ($(2,3)  +(0,-\s)$);
      \coordinate (c\i) at ($(4,0)  +(0,-\s)$);
      \coordinate (d\i) at ($(3,0)  +(0,-\s)$);
      \coordinate (e\i) at ($(2,1.5)+(0,-\s)$);
      \coordinate (f\i) at ($(1,0)  +(0,-\s)$);
    }
    \foreach \i in {1,2} {
      \draw (a\i)--(b\i)--(c\i)--(d\i)--(e\i)--(f\i)--cycle;
    }
    \draw (a3)--(b3)--(c3)--cycle;
    \draw (b1)++(0,-.75) node{$\mathcal{C}_3$};
    \draw (b2) node{$\bullet$} node[right]{$u$}  ++(0,-.75) node{$\mathcal{C}_2$};
    \draw (b3) node{$\bullet$} node[right]{$u'$} ++(0,-1.5) node{$\mathcal{C}_1$};
  \end{tikzpicture}
  \caption{Proof of Proposition~\ref{pro:pump}.}
  \label{fig:context}
\end{figure}

Therefore, the type of the graph generated by $\C_1 \triangleleft \C_2 \triangleleft \dots \triangleleft \C_2$ is $t$, whatever the number of repetitions of $\C_2$.
Finally, since $\C = \C_1 \triangleleft \C_2 \triangleleft \C_3$ has type $t_0$, we have $\Delta_{\C_3}(t)=t_0$ and thus the type of $\C_1 \triangleleft \C_2 \triangleleft \dots \triangleleft \C_2 \triangleleft \C_3$ is always $t_0$.
We conclude since $t_0$ contains $\phi$.
\end{proof}

\section{A reduction from SAT to \textsc{Succinct-}\texorpdfstring{$\phi$}{ϕ}}
\label{sec:sat}

In this section, for any cw-nontrivial sentence $\phi$,
we present a reduction from SAT to \textsc{Succinct-}$\phi$ or to \textsc{Succinct-}$\neg\phi$
(the second alternative being equivalent to reducing UNSAT to \textsc{Succinct-}$\phi$).
That is, from a SAT instance $S$ we explain how to construct a succinct graph representation combining
the pumping from Proposition~\ref{pro:pump}
and the saturating graph from Proposition~\ref{pro:saturating}.
The goal is to establish the following proposition.

\begin{prop}\label{prop:main-reduction}
  Let $k\in\nn$ and $\phi$ be a fixed cw-nontrivial sentence of quantifier rank $q$ and let $S$ be a SAT formula with $n$ variables.
  We assume that $\Omega_q \models \phi$.
  There exists a succinct graph representation $(N,C)$ with size polynomial in $n$ and constructible in polynomial time from $S$, which represents a graph $G$ with the following properties:
  \begin{enumerate}[label={(\roman*)}]
    \item if there exists an assignment of the variables satisfying $S$ (i.e., $S$ is a positive SAT instance), then $G$ is of the form $\Omega_q \oplus H$ where $H$ has cliquewidth $\leq k$;
    \item otherwise, $G$ is the graph generated by the clique expression $\C_1 \triangleleft \C_2 \triangleleft \dots \triangleleft \C_2 \triangleleft \C_3$ with $2^n$ occurrences of $\C_2$, where $\C_1,\C_2,\C_3$ satisfy the conclusions of Proposition~\ref{pro:pump} applied to $\neg \phi$.
  \end{enumerate}
\end{prop}

This shows Theorem~\ref{thm:main}: since $\Omega_q \oplus H \models \phi$ and $\C_1 \triangleleft \C_2 \triangleleft \dots \triangleleft \C_2 \triangleleft \C_3 \not\models \phi$, the above is a valid reduction from SAT assuming $\Omega_q$ models $\phi$; otherwise we obtain a reduction from UNSAT by applying the proposition to $\neg \phi$. Moreover, the reduction always produces graphs of cliquewidth $\leq k$.

Throughout the proof, for readability, we will sometimes identify the clique expression with the graph it generates.
The recoloring associated with a marked clique decomposition is the composition of all recolorings appearing on the path from $\square$ to the root.
Without loss of generality, up to replacing $\C_2$ by $\C_2 \triangleleft \dots \triangleleft \C_2$ (this does not alter the cliquewidth), we may assume that the recoloring of $\C_2$ is idempotent (this is because every element of an finite semigroup has an idempotent power~\cite[Proposition~6.31]{pinBook}).
We heavily rely on idempotence of the recoloring of $\C_2$ to provide a succinct representation of $\C_1 \triangleleft \C_2 \triangleleft \dots \triangleleft \C_2 \triangleleft \C_3$.
This is explained in the proposition below.

\newcommand{\same}{\mathrm{same}}
\newcommand{\suc}{\mathrm{succ}}
\newcommand{\far}{\mathrm{far}}

\begin{prop}\label{prop:10-tables}
  Consider two vertices $v,v'$ in the graph generated by the clique decomposition $\C_1 \triangleleft \C_2 \triangleleft \dots \triangleleft \C_2 \triangleleft \C_3$.
  Assume that $v$ occurs to the left (or in the same copy) as $v'$.
  There are 10 tables $T_{1,1},T_{1,2,\suc},T_{1,2,\far},T_{1,3},T_{2,2,\same},T_{2,2,\suc},T_{2,2,\far},T_{2,3,\far},T_{2,3,\suc}$ and $T_{3,3}$, each of them with constant size, such that whether there is an edge from $v$ to $v'$ and from $v'$ to $v$ can be read from the tables as follows:
  \begin{enumerate}
  \item first, identify if $v$ and $v'$ belong to $\C_1$, to a copy of $\C_2$ or to $\C_3$, and whether they belong to the same copy, to successive copies (the first copy of $\C_2$ is considered to be a successor of $\C_1$, and $\C_3$ is considered to be a successor to the last copy of $\C_2$) or to copies that are further away;
  \item then, choose the corresponding table, and read the entry corresponding to the positions of $v$ and $v'$ within their respective copies.
  \end{enumerate}
\end{prop}

\begin{proof}
  Recall that the adjacency between two vertices in a graph generated by a clique decomposition depends on the label of their least common ancestor, as well as the compositions of the recolorings appearing on the paths from each of the vertices to the least common ancestor.
  In our case, the least common ancestor always belongs to the copy of $v'$ (this is because the clique decomposition has a linear shape, and $v$ occurs to the left of $v'$).
  Moreover, since the recoloring of $\C_2$ is idempotent, it makes no difference whether there is one or more copies of $\C_2$ lying strictly between the copies of $v$ and $v'$.
  The result easily follows from these observations.
\end{proof}

Without loss of generality, up to replacing $\Omega_q$ with $\Omega_q \oplus \dots \oplus \Omega_q$ and $\C_2$ with $\C_2 \triangleleft \dots \triangleleft \C_2$, we may assume that $\Omega_q$ has as many vertices as $\C_2$ has non-$\square$ leaves.
Let $r$ denote this number, which is a constant (i.e., it does not depend on $n$), and let $n_2 = \lceil \log r \rceil$.
Likewise, let $n_1=\lceil \log |\C_1| \rceil$ and $n_3 = \lceil \log |\C_3| \rceil$. 
Let $m= 2+ \max(n + n_2,n_1,n_3)$.

We will construct a Boolean circuit $C$ with $2m$ inputs and one output, which encodes a graph $G$ of size $N=2^n r+|\C_1|+|\C_3|$.
We start with the following technical statement which will allow us to better organize the vertices and their labels by first applying a permutation of $\{1,\dots,2^m\}$.
We omit a proof since it follows easily (though somewhat tediously) from the standard fact that arithmetic operations (including Euclidean division) can be performed using small circuits.
Let us note $j+I=\{j+i\mid i\in I\}$ for any $I\subset\nn$ and $j\in\nn$.

\begin{prop}
  \label{prop:reorg}
  There is a Boolean circuit $C_0$ with $m$ inputs and $m$ outputs and size polynomial in $n$, which computes a map $f:\{1,\dots,2^m\} \to \{1,\dots,2^m\}$ such that (see Figure~\ref{fig:reorg}):
  \begin{itemize}
    \item for every $i \in \{0,\dots, 2^n-1\}$, it holds that $f(ir+\{0,\dots,r-1\}) = i2^{n_2}+\{0,\dots,r-1\}$;
    \item it holds that $f(2^nr+\{0,\dots,|\C_1|-1\}) = 2^{m-1}+\{0,\dots,|\C_1|-1\}$.
    \item it holds that $f(2^nr+|\C_1|+\{0,\dots,|\C_2|-1\}) = 2^{m-1} + 2^{m-2}+\{0,\dots,|\C_2|-1\}$.
  \end{itemize} 
\end{prop}

\begin{figure}
  \centering
  \begin{tikzpicture}
    \def\myfspace{3}
    \foreach \x in {0,\myfspace}
      \draw (\x,0) -- ++(0,-6);
    \foreach [count=\i] \y/\v in {%
      0/0,%
      1/r,%
      2/2\cdot r,%
      3/ ,%
      4/2^n\cdot r,%
      6/2^n\cdot r+|\mathcal{C}_1|,%
      7.5/N=2^n\cdot r+|\mathcal{C}_1|+|\mathcal{C}_3|,%
      12/2^m%
    }{
      \draw (.1,-.5*\y) -- ++(-.2,0) node[left] {\scriptsize $\v$};
      \coordinate (l\i) at (0,-.5*\y);
    }
    \foreach [count=\i] \y/\v in {%
      0/0,%
      1/r,%
      1.4/2^{n_2},%
      2.4/2^{n_2}+r,%
      2.8/2\cdot 2^{n_2},%
      3.8/ ,%
      4.2/ ,%
      5.2/ ,%
      5.4/2^{n+n_2},%
      6.5/2^{m-1},%
      8.5/2^{m-1}+|\mathcal{C}_1|,%
      9/2^{m-1}+2^{m-2},%
      10.5/2^{m-1}+2^{m-2}+|\mathcal{C}_3|,%
      12/2^m%
    }{
      \draw (\myfspace-.1,-.5*\y) -- ++(.2,0) node[right] {\scriptsize $\v$};
      \coordinate (r\i) at (\myfspace,-.5*\y);
    }
    \node at (.5*\myfspace,.5) {$f$};
    \fill[fill=red,opacity=.25] (l1) -- (r1) -- (r2) -- (l2) -- cycle;
    \fill[fill=red,opacity=.25] (l2) -- (r3) -- (r4) -- (l3) -- cycle;
    \fill[fill=red,opacity=.25] (l3) -- (r5) -- (r6) -- (l4) -- cycle;
    \fill[fill=red,opacity=.25] (l4) -- (r7) -- (r8) -- (l5) -- cycle;
    \fill[fill=blue,opacity=.25] (l5) -- (r10) -- (r11) -- (l6) -- cycle;
    \fill[fill=green,opacity=.25] (l6) -- (r12) -- (r13) -- (l7) -- cycle;
  \end{tikzpicture}
  \caption{Illustration of the permutation $f$ in Proposition~\ref{prop:reorg}.}
  \label{fig:reorg}
\end{figure}

We say that a number in $\{1,\dots, 2^m\}$ is \emph{useful} if it is in the image of $f$.
For useful numbers, we may give the following interpretation to the bits in their base-2 decomposition.
The first bit determines whether the number is \emph{principal} (if the bit is 0) or \emph{secondary} (if it is 1).
For principal numbers, the $n_2$ rightmost bits are called \emph{local identifiers}; they encode different vertices within copies of $\C_2$ or $\Omega_q$.
The $n$ next bits are called the \emph{$S$-bits}, and they encode truth values for variables of the SAT formula $S$.
For secondary numbers, the second bit encodes whether we belong to $\C_1$ or $\C_3$, and then the $n_1$ or $n_3$ rightmost bits, accordingly, are again \emph{local identifier} and encode the different vertices of $\C_1$ and $\C_3$.
Here is a summary:
\[
\overbrace{00\dots0 \underbrace{x\dots x}_{S\text{-bits}} \underbrace{\ell \dots \ell}_{\text{local id.}}}^{\text{principal number}} \qquad \overbrace{10\dots0 \underbrace{\ell \dots \ell}_{\text{local id.}}}^{\text{secondary number for } \C_1} \qquad 
\overbrace{110\dots0 \underbrace{\ell \dots \ell}_{\text{local id.}}}^{\text{secondary number for } \C_3}
\]

We are now ready to prove Proposition~\ref{prop:main-reduction}.

\begin{proof}[Proof of Proposition~\ref{prop:main-reduction}]
  Let $u$ and $u'$ be two input numbers in $\{1,\dots,2^m\}$.
  First, we apply circuit $C_0$ to $u$ and $u'$, this yields useful integers $v=f(u)$ and $v'=f(u')$.
  There are a few cases (which are easily distinguished by a Boolean circuit).
  \begin{enumerate}
    \item If either $v$ or $v'$ is a principal number, whose $S$-bits encode a positive SAT assignment, then
    \begin{itemize}
      \item[(1.1)]\label{i:sames} if the other vertex has the same $S$-bits, adjacency is declared by feeding the local identifiers to a predetermined constant size circuit $C_{\Omega_q}$ which encodes the graph $\Omega_q$;
      \item[(1.2)]\label{i:noedge} otherwise, we declare that there is no edge between the vertices.
    \end{itemize}
    \item\label{i:neg} Otherwise, $v$ is either non-principal, in which case it encodes a vertex of $\C_1$ or $\C_3$ (which is decided according to its second bit), or it is principal but its $S$-bits corresponds to a negative SAT assignment, in which case it encodes a vertex of a certain copy of $\C_2$ (more precisely, the $i$-th copy, where $i$ is the binary number corresponding to the $S$-bits). The same is true for $v'$.
    Then we decide whether there is an edge from $v$ to $v'$, following Proposition~\ref{prop:10-tables}, by testing whether $v$ and $v'$ are in the same or in successive copies, and then feeding the local identifiers to one of the 10 predetermined circuits corresponding to the 10 aforementioned tables.
  \end{enumerate}
  Thanks to \eqref{i:sames}, each positive SAT assignment provides a copy of $\Omega_q$, which is disconnected from all other vertices by \eqref{i:noedge}.
  Any other vertex is connected according to item \eqref{i:neg} which, together with Proposition~\ref{prop:reorg}, guarantees that we encode a subgraph $H'$ of the graph generated by the clique expression $\C_1 \triangleleft \C_2 \triangleleft \dots \triangleleft \C_2 \triangleleft \C_3$ on this set of vertices. $H'$ is always a graph of cliquewidth $\leq k$ by \cite[Corollary 3.3]{Courcelle_2000}, hence the entire graph produced by the reduction is always of cliquewidth $\leq k$ (since it is $H'$ plus zero or more disjoint copies of $\Omega_q$).
  Moreover, in the case when there is no positive SAT assignment, $H'$ is exactly the graph produced by the clique expression $\C_1 \triangleleft \C_2 \triangleleft \dots \triangleleft \C_2 \triangleleft \C_3$, since in this case, connectivity between vertices is given by \eqref{i:neg}.
  Proposition~\ref{prop:main-reduction} follows.
\end{proof}

\section{Cw-trivial formulae}
\label{sec:bad}

If $\phi$ is an MSO sentence with finitely many models or with finitely many countermodels, then it can be tested in constant time (recall our convention that Boolean circuits are never bigger than the encoded graph).
We call such a $\phi$ \emph{trivial}, which of course implies that it is cw-trivial.

In this section, we focus on \emph{cw-trivial}, yet \emph{nontrivial} sentences, when neither Theorem~\ref{thm:main}, nor the above remark can be applied.
{On families of graphs with bounded cliquewidth, cw-trivial formulae are answered in constant time.}

For instance, the question ``is $\SGR{N,C}$ a square directed grid?'' is expressible in MSO (see \cite[Proposition 5.14]{courcellebook}),
but is both nontrivial (there are infinitely many square directed grids) and cw-trivial (only finitely many square grids have a cliquewidth bounded by each $k$, a consequence of, for instance, \cite{twcwbpk}).

Even though our theorem does not apply, it is easy to show that this question is \coNP-hard: given an instance $S$ of SAT with $2s$ variables,
construct a circuit for a graph on $N=2^{2s}$ vertices, that views each $n$-bit vertex label as an assignment for $S$, and evaluates it.
If it finds ``false'', then connects the vertices that should be neighbors in the grid according to a fixed natural encoding.
If it finds ``true'', then connects some other neighbors in a way that ensures that the graph will not be a grid (inserting a local triangle for instance).
The succinctly represented graph is a square grid if and only if all assignments for $S$ evaluate to false.
Producing Boolean circuits which evaluate a given instance of SAT is easily done in polynomial time,
which concludes the reduction.
The reader probably see similarities between the above argument and the structure of the reduction from Section~\ref{sec:sat}: an easily computable family of succinct models is combined with a local pattern of counter-model.

It is tempting to conjecture that our main results actually holds in much greater generality, and for instance that for every nontrivial MSO formula $\phi$, the problem \textsc{Succinct-}$\phi$ is either \NP- or \coNP-hard.
We will see in this section that this is very unlikely.
It is nevertheless quite hard to disprove, since for our proof not to apply, any counter-example needs enough irregularity to avoid pumpability (e.g. it has no saturating subgraph).

The construction we present below relies on a complexity hypothesis about the existence of polynomial time algorithms solving hard problems on a large set of instance sizes (but not necessarily on all sizes). We could use the hypothesis ``\textit{there is no polynomial time algorithm solving SAT on an infinite set of sizes of instances}'', but it seems to be a bit strong. Instead, following \cite{fs17}, we will use the concept of \emph{robust set of sizes} which is a notion of ``thickness'' that is robust under applying polynomial time reductions. This allows us to prove our result under a more convincing hypothesis of the kind: ``\textit{there are some problems of the polynomial hierarchy that cannot be solved in polynomial time on an infinite (robust) set of sizes of instances}'' (see Corollary~\ref{cor:bad}).


\begin{defiC}[{\cite[Definition~1]{fs17}}]
  A set $M$ of integers is \emph{robust} if and only if:
  \begin{equation*}
    \forall k, \exists \ell \geq 2: \{\ell,\ell+1,\dots,\ell^k\} \subset M.
  \end{equation*}
  This implies that $M$ is infinite.
\end{defiC}

We denote by UNSAT the complement of SAT (i.e., CNF formulae that are not~\mbox{satisfiable).}  

\begin{thm}
  \label{thm:bad}
  There is a nontrivial first-order sentence $\psi$ such that, if either SAT or UNSAT reduces to \textsc{Succinct-}$\psi$,
  then there is a polynomial-time algorithm solving SAT for a robust set of sizes of instances.
\end{thm}

According to \cite[Proposition~8, Theorem~12]{fs17}, our Theorem~\ref{thm:bad} implies:

\begin{cor}\label{cor:bad}
  Let $\psi$ be given by Theorem~\ref{thm:bad}.
  If either SAT or UNSAT reduces to \textsc{Succinct-}$\psi$,
  then any problem in the polynomial hierarchy can be solved in polynomial time on a robust set of sizes of instances.
\end{cor}

The rest of this section is a proof of Theorem~\ref{thm:bad}.
Our construction relies on previous works from finite model theory dealing with spectrum of FO formulae.
The key idea is that by a diagonalization argument we can build a formula with such a sparse and irregular spectrum that it is sufficiently resistant against polynomial reductions.

For now, let $\psi$ be an arbitrary first-order sentence,
and $f$ a polynomial reduction from either SAT or UNSAT to \textsc{Succinct-}$\psi$.
If $S$ is a SAT (or UNSAT) instance, denote by ${G_{f(S)}}$ the graph succinctly represented by $f(S)$.

\begin{defi}
  Let $P$ be a polynomial and $n$ a positive integer.
  The reduction $f$ is \emph{$P$-meager in $n$} if and only if
  for every \emph{positive} instance $S$ of SAT with size $n$, we have:
  \begin{equation*}
    |G_{f(S)}| \leq P(n).
  \end{equation*}
  The set of values in which $f$ is $P$-meager is called the \emph{$P$-meagerness set} of $f$.
\end{defi}

If $f$ is $P$-meager in some integer $n$, then any SAT instance $S$ of size $n$ such that $G_{f(S)}$ has more than $P(n)$ vertices must be a negative instance.
On the other hand, if $G_{f(S)}$ is smaller than $P(n)$, then we can test $\psi$ on $G_{f(S)}$ in time polynomial in $|G_{f(S)}|$ (the degree of the polynomial being the quantifier rank of $\psi$, recall that $\psi$ is a first-order sentence) and therefore in time polynomial in $n$.
Thus we have the following lemma.

\begin{lem}
  \label{lem:meagersat}
  Assume that $f$ is a reduction as before and let $P$ be a polynomial.
  There is a polynomial-time algorithm that solves SAT on all instance sizes that are in the $P$-meagerness set of $f$.
\end{lem}

\begin{proof}
  Given an instance $S$ of SAT with size $n$, compute the network $f(S)$ in polynomial time;
  call $G$ its transition graph.
  We can compute the size of $G$ from $f(S)$ in polynomial time,
  because we have the list of nodes and the state set of each node.
  If $|G| > P(n)$, then return 'false': if $f$ is not $P$-meager in $n$ then the answer doesn't matter, and if $f$ is $P$-meager in $n$ then it means that $S$ is a negative instance of SAT.
  Otherwise compute $G$ itself, which can be done in polynomial time since $|G| \leq P(n)$,
  and evaluate $\psi$ on $G$, which can also be done in polynomial time (the final answer is negated if reduction $f$ is from UNSAT instead of SAT).
\end{proof}

\begin{lem}
  \label{lem:meagerenum}
  Let $f$ be a reduction as before and $P$ be a polynomial.
  If $f$ has a nonrobust $P$-meagerness set then, for every integer $d\geq 1$, either:
  \begin{enumerate}[label=(\roman*)]
    \item $f$ produces a model of $\psi$ whose size is not in $\nn^{(d)}=\{n^d\mid n\in\nn\}$; or
  \item there exists an increasing primitive-recursive sequence $\mu$ such that $\mu(n)^d$ is the size of a model of $\psi$ for each $n$.
  \end{enumerate}
\end{lem}

\begin{proof}
  By hypothesis there is an integer $k$ such that, for every $\ell \geq 2$,
  there is at least one value among $\ell,\ell+1,\dots,\ell^k$ in which $f$ is not $P$-meager.
  Observe that if $P'$ is the polynomial giving the execution time of $f$,
  then $f$ produces networks whose transition graphs have size at most $2^{P'(n)}$
  with $n$ the size of the SAT or UNSAT instance.
  Let $t:\nn\to\nn$ be a primitive recursive function such that, for every $n$:
  \begin{equation}
    \label{eq:t}
    t(n+1) > \max\{2^{P'(t(n))}, t(n)^k\}.
  \end{equation}
  Moreover since the meagerness set of $f$ is nonrobust,
  we can additionally choose $t$ such that $f$ is not $P$-meager in $t(n)$, for every $n$.
  (Starting from a function $t$ that satisfies Equation~\eqref{eq:t}, given $n$,
  it is possible to enumerate all values $t(n),t(n)+1,\dots,t(n)^k$ and to find the one for which $f$ is not $P$-meager.
  That computation can be done in a primitive recursive fashion.)

  For every $n$, let us pass to $f$ a positive instance of SAT with size $t(n)$
  (or a negative instance if reduction $f$ is from UNSAT instead of SAT);
  the result is a sequence of automata networks, $(\alpha(n))_{n\in\nn}$.
  This sequence can be made primitive recursive because $t$ and $f$ are.
  Call $\beta(n)$ the transition graph of $\alpha(n)$;
  since $f$ is a reduction and we passed positive instances of SAT to it,
  the graph $\beta(n)$ is a model of $\psi$.
  By non-$P$-meagerness of $f$ in $t(n)$, we can furthermore compute
  the positive instances of SAT such that:
  \begin{equation*}
    P(t(n))\leq |\beta(n)|,
  \end{equation*}
  in a primitive recursive fashion
  (simply by enumerating the satisfiable propositional formulae of size $t(n)$).
  Also, considering that $P'$ is the running time of $f$,
  that a transition graph is at most exponential in the size of the network,
  and Equation~\eqref{eq:t}, we have:
  \begin{equation*}
    |\beta(n)|  \leq 2^{P'(t(n))} < t(n+1).
  \end{equation*}
  Since $P(t(n)) \leq |\beta(n)| < P(t(n+1))$, the sequence $|\beta(n)|$ is increasing in $n$.

  Moreover for every $d\geq 1$, there is an $n_0$ such that for every ${n\geq n_0}$,
  we have $|\beta(n+2)|^{1/d} > |\beta(n)|^{1/d}$ (as $|\beta(n+2)|> 2^{P'(|\beta(n)|)}$).
  If ${|\beta(n)|^{1/d}}$ is not an integer for some $n$ then the reduction $f$ produces a model of $\psi$ whose size is not in ${\nn^{(d)}}$ and the lemma follows.
  Otherwise, we define the map $\mu:\nn\to\nn$ by $\mu(n) = |\beta(2(n+n_0))|^{1/d}$;
  by the previous inequality $\mu$ is an increasing map.
  Finally, since $\alpha(n)$ is primitive recursive in $n$, so is $\beta(n)$ and so is $\mu(n)$.
  The lemma follows because by construction ${\mu(n)^d}$ is always the size of a model of $\phi$.
\end{proof}

\begin{defi}
  The \emph{spectrum} of $\psi$ is the set of sizes of finite models of $\psi$.
  In symbols:
  \begin{equation*}
    \spec(\psi) = \{|G|\mid G\models\psi,|G|<\infty\}.
  \end{equation*}
  See~\cite{djmm12} for a survey about this notion.

  A function $h:\nn\to\nn$ is \emph{time-constructible} if and only if
  there is a Turing machine that, for every $n$, halts in exactly $h(n)$ steps on input $n$ written in binary.
\end{defi}

\begin{lem}
  \label{lem:specim}
  For every time-constructible function $h$,
  there exist a first-order sentence $\psi_h$ and an integer $d$ such that $\spec(\psi_h)=\im(h)^{(d)}$, where
  $\im(h)^{(d)} = \{n^d \mid n \in \im(h)\}$.
\end{lem}

\begin{proof}
  By~\cite[Theorem 4.5]{js75} and~\cite[Theorem 3]{f75},
  we just have to prove that $\im(h)$ is a \NEXPTIME{} language.
  Given $n$, the algorithm guesses a word $u$ of length at most $n$,
  runs $h$ on input $u$ and checks that it halts in $n$ steps exactly.
  If $n\in\im(h)$ then there is $u$ such that $h(u)=n$ and $|u|\leq n$ 
  because the machine cannot read more than $n$ input symbols within $n$ steps.
\end{proof}

\begin{lem}
  \label{lem:diag}
  There is a first-order sentence $\psi$ and an integer $d$ such that, for every increasing primitive-recursive function $R$, we have:
  \begin{equation}
    \label{eq:specpsi}
    \im(R)^{(d)} \not\subseteq \spec(\psi)\subseteq\nn^{(d)}.
  \end{equation}
\end{lem}

\begin{proof}
  Let $(R_n)_{n\in\nn}$ be a computable enumeration of increasing primitive-recursive functions.
  (To construct one, start from a computable enumeration of primitive-recursive functions $(R'_n)_{n\in\nn}$
  \cite[Exercice I.7.4]{odifreddi1992classical} and change $R'_n$ into $R_n$ as follows: $R_n:i\mapsto \max\{R_n(i-1)+1, R'_n(i)\}$.
  This transformation is computable and leaves increasing functions unchanged, so it hits all of them.)
  
  For every integer $n$, define the set:
  \begin{equation}
    \label{eq:en}
    E(n) = \{R_i(j) \mid 0\leq i,j\leq n\},
  \end{equation}
  and let $h$ denote an increasing time-constructible function such that:
  \begin{equation}
    \label{eq:hn}
    h(n) > \max E(n),
  \end{equation}
  for instance, $h$ may explicitly compute $\max E(n)$ and spend that many steps idling by decreasing a counter.
  
  Let us show that for every $n$, there exists an element in $\im(R_n)\setminus\im(h)$.
  The set $\{h(0), \dots, h(n-1)\}$ cannot contain $\{R_n(0), \dots, R_n(n)\}$ because $R_n$ is injective
  (since it is increasing).
  So there is an element $i$ of $\{0, \dots, n\}$ such that $R_n(i)$ does not belong to $\{h(0), \dots, h(n-1)\}$.
  We have $R_n(i) < h(n)$ by Equations~\eqref{eq:en}--\eqref{eq:hn};
  since $h$ is increasing, $R_n(i)$ is not in $\im(h)$.
  The existence of the desired formula $\psi$ follows from Lemma~\ref{lem:specim}.
\end{proof}

\begin{proof}[Proof of Theorem~\ref{thm:bad}]
  Let $\psi$ and $d$ be given by Lemma~\ref{lem:diag} and assume that $f$ is a polynomial reduction from either SAT or UNSAT to \textsc{Succinct-}$\psi$.
  Both the spectrum of $\psi$ and its complement are infinite, so $\psi$ is nontrivial.
  If there exists some polynomial $P$ such that $f$ has a robust meagerness set,
  then by Lemma~\ref{lem:meagersat} there exists a polynomial-time algorithm solving SAT on a robust set of instance sizes.
  Otherwise, since $\spec(\psi)\subseteq\nn^{(d)}$, Lemma~\ref{lem:meagerenum} implies that there is an increasing primitive-recursive map $\mu$ such that,
  for every $n$, the quantity $\mu(n)^d$ is the size of a model of $\psi$.
  But then, by Lemma~\ref{lem:diag}, one of those sizes will not be contained in the spectrum of $\psi$: a contradiction.
\end{proof}

\section{Deciding cw-triviality}\label{sec:deciding-cw-triviality}

In this section, we establish two complementary results regarding decidability of cw-triviality

\begin{prop}\label{prop:cw-triviality-undecidable}
  The problem which takes as input an MSO sentence $\phi$ and outputs whether $\phi$ is cw-trivial is undecidable, even if $\phi$ is a first-order sentence.
\end{prop}

\begin{proof}
  Libkin's proof~\cite[Theorem~9.4]{l04} of Trakhtenbrot's theorem~\cite{Trakhtenbrot1950} about undecidability of finite satisfiability of first-order sentences constructs, for any given Turing machine $M$, a first-order sentence $\phi_M$ such that $\phi_M$ admits a model if and only if $M$ halts on the empty input\footnote{In~\cite[Theorem~9.4]{l04}, the author uses a signature with some additional binary relations and a constant symbol. However, this is easily encoded into the signature of graphs, see for instance~\cite[Exercise 9.1]{l04}.}.
  Intuitively, $\phi_M$ corresponds to the property ``the graph is a colored grid encoding a run of the Turing machine $M$''.

  Clearly $\phi_M$ admits infinitely many countermodels of bounded cliquewidth (in fact, almost all graphs are countermodels of $\phi_M$), for instance all paths are countermodels.
  It is easy to modify $\phi_M$ into another first-order sentence $\phi'_M$ such that a graph is a model of $\phi'_M$ if and only if it is the union of a model of $\phi_M$ together with arbitrarily many isolated vertices.
  Again, $\phi'_M$ clearly admits all paths as countermodels.
  
  We claim that $\phi'_M$ admits infinitely many models of cliquewidth $\leq k$, for some $k$, if and only if $M$ halts on the empty input.
  Indeed, if $M$ does not halt, then $\phi_M$ does not have any finite model and thus neither does $\phi'_M$.
  Conversely, if $M$ halts, then $\phi_M$ admits a finite model $G$.
  Let $k$ be the cliquewidth of $G$.
  Then all graphs obtained from $G$ by adding any number of isolated vertices are models of $\phi'_M$ with cliquewidth $\leq k$.
\end{proof}

The next statement follows from a simple pumping argument (see also \cite[Section 1.2]{comon08}) applied to tree automata (see \cite{courcellebook}); we explicit here a more self-contained version.
\begin{prop}\label{prop:cw-triviality-decidable}
  It is decidable, when given an MSO sentence $\phi$ and an integer $k$ as inputs, whether $\phi$ admits infinitely many models of cliquewidth $\leq k$.
\end{prop}
\begin{proof}
  Let $q$ be the quantifier rank of $\phi$.
  Recall from the preliminaries the tree-automaton over $T_q$, which labels clique decompositions of width $k$ by their quantifier rank $q$ types.
  Now note that a type in $T_q$ admits infinitely many models of cliquewidth $\leq k$ if and only if it labels a clique decomposition of depth $d$ with ${|T_q|+1\leq d\leq 2|T_g|}$: indeed, if a type has infinitely many models then it has models of arbitrary depth and one can cut any repetition of a type in a root-to-leaf path of a clique decomposition (see Proposition~\ref{pro:pump} and Figure~\ref{fig:context}) and therefore reduce the depth by at least $1$ and at most $|T_q|-1$ as soon as it greater than ${2|T_q|}$.
  Conversely, if it has a model of depth $>|T_q|$ then it has infinitely many models by pumping (see for instance the proof of Proposition~\ref{pro:pump}).

  To conclude, recall that ${|T_q|}$ can be recursively bounded from $q$ (counting all possible formulae with quantifier rank $\leq q$ regardless of their satisfiability) and that, given $k$ and $d$, there is a finite and computable set of graphs having a clique decomposition of width $\leq k$ and depth $\leq d$. From the discussion above, we therefore have a finite computable list of graphs to test in order to decide cw-triviality of width $k$ of $\phi$.
\end{proof}

\section{Discussion and Future work}

  From the proof of Theorem~\ref{t:succinct}, one can get convinced that the hardness result remains true for even more restricted classes of graphs: for example, when the clique decomposition itself can be succinctly represented, and is very close to a path.
  We could also generalize without difficulty the result from MSO to CMSO, where the signature is extented to include predicates for cardinality of sets modulo some integer.
  This is because this extension does not make the number of types grow infinite.
  Actually, the proofs could work for classes, further than MSO, that involve saturation and tools to glue.

One could consider different parametrizations of \textsc{Succinct-}$\phi$.
Let us first point out that parametrization by the size or quantification rank of the formulae fails.
The formula $\exists x: E(x,x)$ expresses the property: ``the graph has at least one loop''
(or the transition graph has at least one fixed point, in terms of automata networks).
That question is \NP-hard: given an instance $S$ of SAT, produce a succinct graph representation
$(2^{|S|},C)$ with labels on $|S|$ bits that evaluates it on $S$;
if it finds ``true'', then the only edge is to itself; if it finds ``false'', then the only edge is to the lexicographic next label (cyclically).
The formula above is virtually the smallest possible, according to to any reasonable parameterization of logic formulae,
so there is no hope to get fixed-parameter tractability if the parameter concerns the formula.
Moreover, as pointed in Theorem~\ref{thm:main}, 
parameterization by the cliquewidth of graphs also fails:
we have \NP- or \coNP-hardness even when the cliquewidth is guaranteed to be at most~$2$.

Another relevant parameter, when taking the automata networks point of view, is the size of the alphabets $Q_v$ used at each automaton.
The natural requirement that all automata hold the same alphabet $Q$ introduces additional arithmetical constraints (the number of vertices must be a power of $|Q|$).
For example, a cw-nontrivial formula asking that each configuration belong to a cycle of length $2$ becomes trivial for ternary-alphabet automata networks.
This point of view is adopted in \cite{rice-ban}.

On the other hand, it remains to fully characterize which cw-trivial MSO sentences yield an \NP- or \coNP-hard problem,
and what happens with those that do not.
In particular, Theorem~\ref{thm:bad} does not say whether the formula $\phi$ has a polynomial-time solvable \textsc{Succinct-}$\phi$ problem.
We do not know whether it could be the case for some $\phi$ under reasonable complexity assumptions.
From our construction, we can nevertheless deduce that a cw-nontrivial MSO sentence $\phi$
is either proven to be \NP-hard when $\Omega_q\models\phi$,
or to be \coNP-hard when $\Omega_q\not\models\phi$.

\section*{Acknowledgments}
\noindent The authors are very grateful to Édouard Bonnet for a proof of Proposition~\ref{pro:saturating}, and Colin Geniet for fruitful discussions about (un-)decidability within our settings.

\bibliographystyle{alphaurl}
\bibliography{ricean-mso-nd.bib}

\newcommand{\etalchar}[1]{$^{#1}$}
\begin{thebibliography}{DJMM12}

\bibitem[Ara08]{a08}
J.~Aracena.
\newblock Maximum number of fixed points in regulatory boolean networks.
\newblock {\em Bulletin of Mathematical Biology}, 70:1398--1409, 2008.
\newblock \href {https://doi.org/10.1007/s11538-008-9304-7} {\path{doi:10.1007/s11538-008-9304-7}}.

\bibitem[BLT92]{Balcazar_1992}
J.~L. Balc{\'a}zar, A.~Lozano, and J.~Tor{\'a}n.
\newblock The complexity of algorithmic problems on succinct instances.
\newblock {\em Computer Science}, pages 351--377, 1992.
\newblock \href {https://doi.org/10.5555/166961.167022} {\path{doi:10.5555/166961.167022}}.

\bibitem[CDG{\etalchar{+}}08]{comon08}
H.~Comon, M.~Dauchet, R.~Gilleron, F.~Jacquemard, D.~Lugiez, C.~L{\"o}ding, S.~Tison, and M.~Tommasi.
\newblock {\em Tree Automata Techniques and Applications}.
\newblock 2008.
\newblock URL: \url{https://hal.science/hal-03367725}.

\bibitem[CE12]{courcellebook}
B.~Courcelle and J.~Engelfriet.
\newblock {\em Graph structure and monadic second-order logic. A language-theoretic approach.}
\newblock Encyclopedia of Mathematics and its applications, Vol. 138. {C}ambridge University Press, 2012.
\newblock \href {https://doi.org/10.1017/CBO9780511977619} {\path{doi:10.1017/CBO9780511977619}}.

\bibitem[CMR00]{courcelle-cw}
B.~Courcelle, J.~Makowsky, and U.~Rotics.
\newblock Linear time solvable optimization problems on graphs of bounded clique-width.
\newblock {\em Theory of Computing Systems}, 33:125–--150, 2000.
\newblock \href {https://doi.org/10.1007/s002249910009} {\path{doi:10.1007/s002249910009}}.

\bibitem[CO00]{Courcelle_2000}
B.~Courcelle and S.~Olariu.
\newblock Upper bounds to the clique width of graphs.
\newblock {\em Discrete Applied Mathematics}, 101(1-3):77--114, 2000.
\newblock \href {https://doi.org/10.1016/s0166-218x(99)00184-5} {\path{doi:10.1016/s0166-218x(99)00184-5}}.

\bibitem[Cul71]{c71}
P.~Cull.
\newblock Linear analysis of switching nets.
\newblock {\em Biological Cybernetics}, 8:31--39, 1971.
\newblock \href {https://doi.org/10.1007/BF00270831} {\path{doi:10.1007/BF00270831}}.

\bibitem[DF13]{df13}
R.~Downey and M.~Fellows.
\newblock {\em Fundamentals of Parameterized Complexity}.
\newblock Springer Verlag London, 2013.
\newblock \href {https://doi.org/10.1007/978-1-4471-5559-1} {\path{doi:10.1007/978-1-4471-5559-1}}.

\bibitem[DJMM12]{djmm12}
A.~Durand, N.~D. Jones, J.~A. Makowsky, and M.~More.
\newblock Fifty years of the spectrum problem: survey and new results.
\newblock {\em Bulletin of Symbolic Logic}, 18(4):505--553, 2012.
\newblock \href {https://doi.org/10.2178/bsl.1804020} {\path{doi:10.2178/bsl.1804020}}.

\bibitem[DS20]{ds20}
J.~Demongeot and S.~Sen{\'{e}}.
\newblock About block-parallel boolean networks: a position paper.
\newblock {\em Natural Computing}, 19(1):5--13, 2020.
\newblock \href {https://doi.org/10.1007/s11047-019-09779-x} {\path{doi:10.1007/s11047-019-09779-x}}.

\bibitem[DST17]{DAS2017436}
B.~Das, P.~Scharpfenecker, and J.~Torán.
\newblock Cnf and dnf succinct graph encodings.
\newblock {\em Information and Computation}, 253:436--447, 2017.
\newblock \href {https://doi.org/10.1016/j.ic.2016.06.009} {\path{doi:10.1016/j.ic.2016.06.009}}.

\bibitem[Els59]{e59}
B.~Elspas.
\newblock The theory of autonomous linear sequential networks.
\newblock {\em IRE Transactions on Circuit Theory}, 6:45--60, 1959.
\newblock \href {https://doi.org/10.1109/TCT.1959.1086506} {\path{doi:10.1109/TCT.1959.1086506}}.

\bibitem[Fag75]{f75}
R.~Fagin.
\newblock A spectrum hierarchy.
\newblock {\em Zeitschrift für mathematische Logik und Grundlagen der Mathematik}, 21:123--134, 1975.
\newblock \href {https://doi.org/10.1002/malq.19750210117} {\path{doi:10.1002/malq.19750210117}}.

\bibitem[FRRS09]{frrs09}
M.~R. Fellows, F.~A. Rosamond, U.~Rotics, and S.~Szeider.
\newblock Clique-width is np-complete.
\newblock {\em SIAM Journal on Discrete Mathematics}, 23(2):909--939, 2009.
\newblock \href {https://doi.org/10.1137/070687256} {\path{doi:10.1137/070687256}}.

\bibitem[FS17]{fs17}
L.~Fortnow and R.~Santhanam.
\newblock Robust simulations and significant separations.
\newblock {\em Information and Computation}, 256:149--159, 2017.
\newblock \href {https://doi.org/10.1016/j.ic.2017.07.002} {\path{doi:10.1016/j.ic.2017.07.002}}.

\bibitem[GGPT21]{ggpt21}
G.~Gamard, P.~Guillon, K.~Perrot, and G.~Theyssier.
\newblock Rice-like theorems for automata networks.
\newblock In {\em Proceedings of STACS'2021}, volume 187 of {\em LIPICs}, pages pp.\ 8:1--8:15, 2021.
\newblock \href {https://doi.org/10.4230/LIPIcs.STACS.2021.32} {\path{doi:10.4230/LIPIcs.STACS.2021.32}}.

\bibitem[GLP24]{rice-ban}
A.~Goubault-Larrecq and K.~Perrot.
\newblock {R}ice-like complexity lower bounds for {B}oolean and uniform automata networks, 2024.
\newblock Preprint.
\newblock \href {https://arxiv.org/abs/2409.08762v1} {\path{arXiv:2409.08762v1}}, \href {https://doi.org/10.48550/arXiv.2409.08762} {\path{doi:10.48550/arXiv.2409.08762}}.

\bibitem[GM90]{gm90}
E.~Goles and S.~Martinez.
\newblock {\em Neural and Automata Networks: Dynamical Behavior and Applications}.
\newblock Kluwer Academic Publishers, 1990.
\newblock \href {https://doi.org/10.1007/978-94-009-0529-0} {\path{doi:10.1007/978-94-009-0529-0}}.

\bibitem[GW83]{Galperin_1983}
H.~Galperin and A.~Wigderson.
\newblock Succinct representations of graphs.
\newblock {\em Information and Control}, 56(3):183--198, mar 1983.
\newblock \href {https://doi.org/10.1016/s0019-9958(83)80004-7} {\path{doi:10.1016/s0019-9958(83)80004-7}}.

\bibitem[GW00]{twcwbpk}
F.~Gurski and E.~Wanke.
\newblock {\em The Tree-Width of Clique-Width Bounded Graphs without Kn,n}, pages 196--205.
\newblock Springer Berlin Heidelberg, 2000.
\newblock \href {https://doi.org/10.1007/3-540-40064-8_19} {\path{doi:10.1007/3-540-40064-8_19}}.

\bibitem[JS74]{js75}
N.~Jones and A.~Selman.
\newblock Turing machines and the spectra of first-order formulas.
\newblock {\em Journal of Symbolic Logic}, 39(1):139--150, 1974.
\newblock \href {https://doi.org/10.1145/800152.804909} {\path{doi:10.1145/800152.804909}}.

\bibitem[Kau69]{k69}
S.~A. Kauffman.
\newblock Metabolic stability and epigenesis in randomly constructed genetic nets.
\newblock {\em Journal of Theoretical Biology}, 22:437--467, 1969.
\newblock \href {https://doi.org/10.1016/0022-5193(69)90015-0} {\path{doi:10.1016/0022-5193(69)90015-0}}.

\bibitem[KS08]{ks08}
G.~Karlebach and R.~Shamir.
\newblock Modelling and analysis of gene regulatory networks.
\newblock {\em Nature Reviews Molecular Cell Biology}, 9:770--780, 2008.
\newblock \href {https://doi.org/10.1038/nrm2503} {\path{doi:10.1038/nrm2503}}.

\bibitem[Lib04]{l04}
L.~Libkin.
\newblock {\em Elements of Finite Model Theory}.
\newblock Springer Berlin Heidelberg, 2004.
\newblock \href {https://doi.org/10.1007/978-3-662-07003-1} {\path{doi:10.1007/978-3-662-07003-1}}.

\bibitem[MAB98]{ma98}
L.~Mendoza and E.~R. Alvarez-Buylla.
\newblock Dynamics of the genetic regulatory network for \emph{Arabidopsis thaliana} flower morphogenesis.
\newblock {\em Journal of Theoretical Biology}, 193:307--319, 1998.
\newblock \href {https://doi.org/10.1006/jtbi.1998.0701} {\path{doi:10.1006/jtbi.1998.0701}}.

\bibitem[Odi92]{odifreddi1992classical}
P.~Odifreddi.
\newblock {\em Classical Recursion Theory}.
\newblock Elsevier, 1992.

\bibitem[Pin25]{pinBook}
J.-{\'E}. Pin.
\newblock {\em Mathematical Foundations of Automata Theory}.
\newblock 2025.
\newblock Lecture notes from MPRI course (available online).
\newblock URL: \url{https://www.irif.fr/~jep/PDF/MPRI/MPRI.pdf}.

\bibitem[PY86]{Papadimitriou_1986}
C.~H. Papadimitriou and M.~Yannakakis.
\newblock A note on succinct representations of graphs.
\newblock {\em Information and Control}, 71(3):181--185, dec 1986.
\newblock \href {https://doi.org/10.1016/s0019-9958(86)80009-2} {\path{doi:10.1016/s0019-9958(86)80009-2}}.

\bibitem[Rob86]{r86}
F.~Robert.
\newblock {\em Discrete Iterations: A Metric Study}.
\newblock Springer Verlag, 1986.
\newblock \href {https://doi.org/10.1007/978-3-642-61607-5} {\path{doi:10.1007/978-3-642-61607-5}}.

\bibitem[Tho73]{t73}
R.~Thomas.
\newblock Boolean formalization of genetic control circuits.
\newblock {\em Journal of Theoretical Biology}, 42:563--585, 1973.
\newblock \href {https://doi.org/10.1016/0022-5193(73)90247-6} {\path{doi:10.1016/0022-5193(73)90247-6}}.

\bibitem[Tra50]{Trakhtenbrot1950}
B.~A. Trakhtenbrot.
\newblock The impossibility of an algorithm for the decidability problem on finite classes.
\newblock {\em Doklady Akademii Nauk SSSR}, 70(4):569--572, 1950.
\newblock In Russian.

\end{thebibliography}

\end{document}